\magnification=\magstephalf
\overfullrule=0pt
\setbox0=\hbox{{\cal W}}
\def\ww{{\cal W}\hskip-\wd0\hskip 0.2 true pt{\cal W}}
\def\w{{\cal W}}
\def\n{{\cal N}}

\def\lb{\lbrack}
\def\rb{\rbrack}

\def\q#1{\lb#1\rb}
\def\mn{\medskip\smallskip\noindent}
\def\sn{\smallskip\noindent}
\def\bn{\bigskip\noindent}

\font\extra=cmss10 scaled \magstep0 \font\extras=cmss10 scaled 750

\setbox1 = \hbox{{{\extra R}}}
\setbox2 = \hbox{{{\extra I}}}
\setbox3 = \hbox{{{\extra C}}}

\setbox4=\hbox{{{\extra Z}}}
\setbox5=\hbox{{{\extras Z}}}
\setbox6=\hbox{{{\extras z}}}

\def\z{{{\extras Z}}\hskip-\wd5\hskip 2 true pt{{\extras Z}}}
\def\zs{{{\extras z}}\hskip-\wd6\hskip 1.7 true pt{{\extras z}}}

\def\zed{\hbox{{\extras\z}}}
\def\zeds{\hbox{{\extras\zs}}}

\def\vac{\mid \!v \rangle\, }


\font\bigwfont=cmsy10 scaled \magstep3
\setbox7=\hbox{{\bigwfont W}}
\def\WW{{\bigwfont W}\hskip-\wd7\hskip 0.8 true pt{\bigwfont W}}
\def\section#1{\leftline{\bf #1}
\vskip-7pt
\line{\hrulefill}}
\def\cowww{C_{W W}^W}
\def\and{\phantom{l}\wedge\phantom{l}}
\def\ut{\tilde{u}}
\def\vt{\tilde{v}}
\def\bpz{1}
\def\zam{2}
\def\bbss{3}
\def\blg{4}
\def\drinso{5}
\def\feifre{6}
\def\bouwschou{7}
\def\laszlo{8}
\def\ragoucy{9}
\def\blm{10}
\def\kau{11}
\def\mfl{12}
\def\rva{13}
\def\wirrep{14}
\def\cap{15}
\def\wowoA{16}
\def\wowoB{17}
\def\gin{18}
\def\kri{19}
\def\kausch{20}
\def\flohkau{21}
\def\hornfeck{22}
\def\bouwknegt{23}
\def\kauW{24}
\def\commute{25}
\def\kliem{26}
\def\twists{27}
\def\wattsBn{28}
\def\ahnA{29}
\def\ahnB{30}
\def\zhanghuang{31}
\def\ito{32}
\def\watts{33}
\def\frenkel{34}
\def\horst{35}
\def\bowwatts{36}
\def\schellekens{37}
\def\flm{38}
\def\tuite{39}
\def\dgm{40}
\font\HUGE=cmbx12 scaled \magstep4
\font\Huge=cmbx10 scaled \magstep4
\font\Large=cmr12 scaled \magstep3

\font\large=cmr17 scaled \magstep0
%
%
\nopagenumbers
\pageno = 0
\centerline{\HUGE Universit\"at Bonn}
\vskip 10pt
\centerline{\Huge Physikalisches Institut}
\vskip 2.3cm
\centerline{\Large How Complete is the}
\vskip 6pt
\centerline{\Large Classification of \WW-Symmetries ?}
\vskip 0.8cm
\centerline{by}
\vskip 0.8cm
\centerline{\large W.\ Eholzer, A.\ Honecker, R.\ H\"ubel}
\vskip 0.9cm
\centerline{This is a revised and extended version of a letter
which has already appeared in}
\centerline{Phys.\ Lett.\ {\bf B308} (1993) p.\ 42}
\vskip 0.9cm
\centerline{\bf Abstract}
\vskip 15pt
\noindent
In 2D conformal quantum field theory, we continue a systematic study
of $\w$-algebras with two and three generators and their highest
weight representations focussing mainly on rational models.
We review the known facts about rational models of
$\w(2,\delta)$-algebras.
Our new rational models of $\w$-algebras with two generators all
belong to one of the known series. The majority of $\w$-algebras
with three generators --including the new ones constructed in this
letter-- can be explained as subalgebras
or truncations
of Casimir algebras. Nonetheless, for one solution of $\w(2,4,6)$
we reveal some features that do not fit into the pattern
of Casimir algebras or orbifolds thereof. This shows
that there are more $\w$-algebras than those predicted from
Casimir algebras (or Toda field theories). However, most of
the known rational conformal field theories
belong to the minimal series of some Casimir algebra.
\vfill
\settabs \+&  \hskip 120mm & \phantom{XXXXXXXXXXX} & \cr
\+ & Post address:                       & BONN-HE-93-08      & \cr
\+ & Nu{\ss}allee 12                     & hep-th/9302124     & \cr
\+ & W-5300 Bonn 1                       & Bonn University    & \cr
\+ & Germany                             & February 1993      & \cr
\+ & e-mail:                             & revised September '93 & \cr
\+ & unp06b@ibm.rhrz.uni-bonn.de         & ISSN-0172-8733     & \cr
\eject
\pageno=1
\footline{\hss\tenrm\folio\hss}
%
\section{1.\ Introduction}
\mn
The classification of all conformal field theories which
are rational with respect to some extended conformal
algebra (or `$\w$'-algebra) is an open problem
since rational conformal field theories (RCFTs)
$\q{\bpz}$ and $\w$-algebras $\q{\zam}$ have been discovered.
\sn
For the construction of $\w$-algebras existing for
generic value of the central charge $c$ several methods
have been developed: Casimir constructions $\q{\bbss}$,
conserved currents of quantum Toda Field theories $\q{\blg}$
and quantized Drinfeld-Sokolov reductions $\q{\drinso} \q{\feifre}$
(for a recent review see $\q{\bouwschou}$ and references
therein). All these constructions seem to be
equivalent and there has been some progress classifying
their classical analoga $\q{\laszlo} \q{\ragoucy}$
using group theoretical methods.
\sn
Direct constructions of $\w$-algebras with given
dimensions of the simple fields $\q{\blm} \q{\kau}$
revealed the existence of $\w$-algebras
which cannot be obtained directly by the above constructions.
They appeared for isolated values of the central charge
and the study of their highest weight representations (HWRs)
$\q{\rva}\q{\wirrep}$ provided at least one series of RCFTs $\q{\mfl}$
where the symmetry algebra seems to be no subalgebra
or truncation of some
generically existing $\w$-algebra. However, the majority
of rational models are related to a generically
existing algebra $\q{\bouwschou}$: Some explicitly known models
are minimal models of Casimir algebras
\footnote{${}^{1})$}{\sevenrm
In the earlier version of this letter certain
RCFTs had not been identified as minimal models of Casimir algebras.
}, others are
extensions of the Virasoro algebra and diagonalize the
non-diagonal modular invariant partition functions of
the ADE-classification $\q{\cap} \q{\rva}$.
\sn
Although the rational model of $\w(2,8)$ at $c=-{712 \over 7}$
belongs to the minimal series of the Casimir algebra ${\cal WE}_8$
also a strange construction for its characters
by a projection of a direct sum of tensor products of
RCFTs is known $\q{\wirrep}$. Recently, for
$\w(2,8)$ at $c=-{3164 \over 23}$ an exceptional relation to
representations of the modular group has been found $\q{\wowoA}$
leading to more easily computable expressions for the characters
$\q{\wowoB}$ than those of the corresponding Casimir algebra ${\cal WE}_7$.
$\w(2, \delta)$-algebras also
exist for $c=1-8 \delta$ and $c=1-3 \delta$ with effective
central charge $\tilde{c} = c - 24 h_{{\rm min}}= 1$.
These models completed
the classification of $\tilde{c} = 1$ RCFTs $\q{\mfl}$.
Already the $c=1$-classification of RCFTs $\q{\gin}\q{\kri}$
itself leads to $\w$-algebras. It predicts a $\w(2,4,\delta)$
for arbitrary dimension $\delta$ and the algebras $\w(2,16)$,
$\w(2,9,16)$, $\w(2,36)$ at $c=1$.
Finally, $\w(2, \delta)$-algebras exist for Virasoro
non-minimal values of $c=c_{1,s}$ $\q{\kausch}$.
We verified explicitly that adding two further fields of
dimension $\delta$ restricts the relevant HWRs to a finite set for
$\delta \in \{3,5\}$. The explicit form of the characters
of the corresponding RCFTs is still not completely clear
$\q{\flohkau}$.
\sn
In this letter we construct further $\w(2,\delta)$-algebras
and show that they all belong to one of the above series.
Thus, there is a chance that at least all rational models
of $\w(2,\delta)$-algebras are now known in principle.
\sn
For $\w$-algebras with three and more generators less
is known although direct computations have recently been pushed
forward $\q{\hornfeck}$. One aim of this letter is to continue
a systematic study in order to understand rational models of
$\w$-algebras with three generators better. In particular, for
$\w(2,4,6)$ there is one solution in addition to
the Casimir algebra of ${\cal B}_3 / {\cal C}_3$
and the bosonic projection of the $N=1$ Super Virasoro
algebra $\q{\bouwknegt}$ which has not been explained
yet $\q{\kauW}$. Our study of the HWRs of this solution
shows that it is neither related to a Casimir algebra nor
an orbifold.
\sn
We will use the notations and methods described in
$\q{\blm} \q{\rva} \q{\wirrep} \q{\commute}$.
\mn
\section{2.\ New $\ww$-algebras with two generators}
\mn
First, we discuss new $\w$-algebras with one additional
generator. We denote the additional simple field by
`$W$' and the corresponding dimension by $\delta$.
For $9 \le \delta \le 11$ the Jacobi identities
lead to values of $c$ contained in the following table:
\mn
\centerline{\vbox{
\hbox{
\vrule \hskip 1pt
\vbox{ \offinterlineskip
\def\tablespace{height2pt&\omit&&\omit&&\omit&&\omit&\cr}
\def\tablerule{ \tablespace
                \noalign{\hrule}
                \tablespace        }
\hrule
\halign{&\vrule#&
  \strut\hskip 4pt\hfil#\hfil\hskip 4pt\cr
\tablespace
\tablespace
& $\delta$ &&      $c$           && $\left( C_{W W}^W \right)^2$
                     && {\it interpretation} &\cr
\tablespace
\tablerule
& $9$      && $-{14 \over 11}$   && $0$ && $c=c_{6, 11}$ & \cr \tablespace
& \omit    && $-{1206 \over 19}$ && $0$ && $c=c_{3, 38}$ & \cr \tablespace
& \omit    && $-{208 \over 35}$  && $0$ && $c=c_{5, 14}$ & \cr \tablespace
& \omit    && $-{91 \over 5}$    && $0$ && $c=c_{1,5}$ \quad `$(1,s)$'
                                              & \cr \tablespace
& \omit    && $-71$              && $0$ && $c=1 - 8 \cdot 9$ \quad `parabolic'
                                              & \cr \tablerule
& $10$     && ${8 \over 35}$     && $0$ && $c=c_{7,10}$ & \cr \tablespace
& \omit    && ${25 \over 26}$    && ${35685578698267410 \over 2806248908489}$
                                        && $c=c_{12,13}$ & \cr \tablespace
& \omit    && $-29$              && ${1865959562738287104 \over 9178198719145}$
                                        && $c=1 - 3 \cdot 10$ \quad `parabolic'
                                                              & \cr \tablespace
& \omit    && $-2$               && $-{352836 \over 5}$
                                        && $c=c_{1,2}$
                                  \quad `$(1,s)$'--subalgebra & \cr \tablerule
& $11$     && $-{36 \over 13}$   && $0$ && $c=c_{6, 13}$ & \cr \tablespace
& \omit    && $-{1826 \over 23}$ && $0$ && $c=c_{3, 46}$ & \cr \tablespace
& \omit    && $-24$              && $0$ && $c=c_{1,6}$ \quad `$(1,s)$'
                                                              & \cr \tablespace
}
\hrule}\hskip 1pt \vrule}
\hbox{\quad Table 1: $c$-values and coupling constants for
                     $\w(2,\delta)$-algebras}}
}
\mn
The column `interpretation' contains the parametrization of the central
charge if the $\w$-algebra is either related to the ADE-classification of
modular invariant partition functions $\q{\cap}$, or one of the parabolic
algebras with $c=1-8 \delta$, $c=1-3 \delta$ $\q{\mfl}$
or is related to non-minimal models at $c=c_{1,s}$ $\q{\kausch}$.
\medskip
Now we are going to discuss our results for the individual algebras in more
detail.
\mn
{\bf $\bf \ww(2,9)$, $\bf \ww(2,11):$}
The first algebra has already been studied in $\q{\kliem}$.
The space spanned by the Jacobi polynomials has dimension 2.
In the case of $\w(2,11)$ the space spanned by the Jacobi polynomials is
three dimensional. For both algebras only $\cowww = 0$ is possible.
$\w(2,9)$ is consistent for five values of the central charge.
\sn
For $\w(2,11)$  we are able to check two out of the three conditions if we
use only fields up to dimension 18. Both conditions are satisfied
for the three rational values of the central charge listed in the above
table.
\bn
{\bf $\bf \ww(2,10):$}
The space spanned by the Jacobi polynomials has dimension 3.
All three conditions are satisfied for the four values of $c$
given in the above table.
A null field of dimension $18$ that is linear in $W$ exists for $c=-2$.
In this case $\cowww$ is non-zero so that it is important to notice that
the coupling constant to this null field vanishes. Thus, the algebra is also
consistent for $c=-2$.
\mn
For $\w(2,10)$ with $\cowww = 0$ A.\ Kliem has already derived the
necessary condition $c={8 \over 35}$ $\q{\kliem}$.
Our calculation shows that $\w(2,10)$ is indeed consistent for this value
of the central charge. For $c={8 \over 35}$ and $c={25 \over 26}$
both models are Virasoro-minimal and
the field $W$ can be identified with the field of dimension
$h_{7,10;1,6}=10$ resp.\ $h_{12,13;1,7}=10$ in the Virasoro-minimal
model.
\mn
At $c=-2$ we have $h_{1,2;5,1} = 10$. Therefore $\w(2,10)$ might
be a subalgebra of $\w(2,3)$ for this value of the central charge.
In order to investigate this further, we apply the
Jacobi-identity-method $\q{\rva} \q{\wirrep}$ to the HWRs of
$\w(2,10)$ at $c=-2$. One obtains that one must either have
$$h \in \left\{3,
-{3 \over 32}, {5 \over 32}, {21 \over 32}, {45 \over 32}, {8 \over 15}
\right\} \eqno({\rm 1})$$
with corresponding fixed non-zero $W_0$-eigenvalue $w$
or the following relation among $h$ and $w$ must be satisfied:
$$w = - {\cowww \over 5613300} (8h+1) (8h-3) (h-1) h^2.
 \eqno({\rm 2})$$
Among (1) we do indeed see $h=3$ which should stem from the spin three
field in the original symmetry algebra. (2) is very much reminiscent
of the corresponding relation for $w^2$ in the normal sector of $\w(2,3)$.
Note that the linearity of (2) designates that the field $W$
of the algebra $\w(2,10)$ is quadratic in the spin three field of
$\w(2,3)$. The remaining rational $h$-values in (1) stem from
the twisted sector of $\w(2,3)$. The additional $h$-values in (1)
and the additional zeros in (2) have half-integer difference to those
turning up for $\w(2,3)$ and are thus easily explained by a splitting
of characters.
\sn
The most interesting point in these observations is that our explicit
calculations for $\w(2,10)$ do so neatly reproduce the results for
$\w(2,3)$ at $c=-2$ $\q{\twists}$. We conclude that the $(1,s)$-models are not
rational with respect to the corresponding $\w(2,\delta)$-algebra.
However, these algebras may be considered as subalgebras
of $\w(2,\delta,\delta,\delta)$-algebras $\q{\kausch}$. We have
studied the representations of the larger algebras for
$\delta \in \{3,5\}$ explicitly, revealing rational models
at $c_{1,s}$.
The zero modes of the additional simple fields form representations
of $su(2)$ in the $L_0$-eigenspace to eigenvalue $h$.
In particular, for $\w(2,3,3,3)$ at $c=-2$ the only admissable
representations are singlets with $h \in \{0, -{1 \over 8} \}$ and
doublets with $h \in \{{3 \over 8}, 1 \}$.
Note that this are precisely the zeroes of (2).
\bigskip
Note that the subalgebra-structure at $c=-2$ for $\w(2,3)$ generalizes
to the whole $(1,s)$-series. In order to see this consider the fusion rules
of the fields $\phi_{1,p}$ with dimension $h_{1,s;1,p}$ ($p$ odd)
at $c=c_{1,s}$:
$$\lb \phi_{1,1} \rb \times \lb \phi_{1,p} \rb = \lb \phi_{1,p} \rb \ ,
\qquad \qquad
\lb \phi_{1,p} \rb \times \lb \phi_{1,p} \rb
  = \sum_{{q=1 \atop q \in 2 \zeds + 1}}^{2 p - 1} \lb \phi_{1,q} \rb.
 \eqno({\rm 3})$$
$h_{1,s;1,3} = 2 s - 1$ is always odd,
$h_{1,s;1,5} = 3 h_{1,s;1,3}+1 = 2(3 s - 1)$
always even and one has $h_{1,s;1,7} = 2 h_{1,s;1,5}+1$.
Therefore the fusion rules (3) specialize for $p=3$ and $p=5$ to:
$$\lb \phi_{1,3} \rb \times \lb \phi_{1,3} \rb
    = \lb \phi_{1,1} \rb \ ,
\qquad \qquad
\lb \phi_{1,5} \rb \times \lb \phi_{1,5} \rb
    = \lb \phi_{1,1} \rb + \lb \phi_{1,5} \rb.
 \eqno({\rm 4})$$
Thus, the energy-momentum tensor $L$ (corresponding to $\phi_{1,1}$)
forms a closed subalgebra with the primary field $W = \phi_{1,p}$ for $p=3$
and $p=5$. So, this argument predicts a $\w(2,h_{1,s;1,3})$ and
a $\w(2,h_{1,s;1,5})$ at $c=c_{1,s}$. The next algebras in the second
series should be $\w(2,16)$ at $c=-7$ and $\w(2,22)$ at $c=-{25 \over 2}$.
\bigskip
The construction of $\w(2,9)$ involves fields up to dimension 16.
For $\w(2,10)$ one needs fields up to dimension 18.
$\w(2,12)$ and $\w(2,13)$ are determined by three resp.\ four independent
conditions from the Jacobi identities. As there are major technical problems
dealing even with fields of dimension 18 which would yield only one
condition we did not study these algebras.
\bn
\section{3.\ Mixed $\ww$-algebras with three generators}
\mn
$\w(2,\delta_1,\delta_2)$-algebras that contain bosonic
as well as fermionic fields have not been studied very intensely.
We shall therefore focus on some of these `mixed' algebras.
Let $U$ be the fermionic
field and $V$ the boson. If the dimension of $V$ is odd all coupling
constants vanish. In this case a Jacobi identity of type $(U,U,V)$ cannot
be satisfied and the algebra does not exist (this generalizes the observation
of $\q{\blm}$ for two additional bosonic fields with odd dimension). Thus,
the only interesting cases are those where the dimension of $V$ is even.
For the first few algebras with $\delta_1= d(U) > {3 \over 2}$
and $\delta_2 = d(V) > 2$ we
obtain the following explicit results:
\eject
\centerline{\vbox{
\hbox{
\vrule \hskip 1pt
\vbox{ \offinterlineskip
\def\tablespace{height2pt&\omit&&\omit&&\omit&&\omit&&\omit&&\omit&\cr}
\def\tablerule{ \tablespace
                \noalign{\hrule}
                \tablespace        }
\hrule
\halign{&\vrule#&
  \strut\hskip 4pt\hfil#\hfil\hskip 4pt\cr
\tablespace
\tablespace
& $\delta_1, \delta_2$ && $c$                 && $(C_{UU}^V)^2$
     && $(C_{VV}^V)^2$   && $C_{UU}^V C_{VV}^V$  && {\it interpretation} &\cr
\tablespace
\tablerule
& ${5 \over 2}, 4$ && {\it generic}
                       && ${6 (14 c+13) \over (5 c+22)}$
                          && ${54(2c^2+83c-490)^2 \over
                                       (5 c+22)(14c+13)(2c+25)^2}$
                               && ${18(2c^2+83c-490) \over (5 c+22)(2c+25)}$
                                 && $n = 2$ &\cr \tablerule
& ${7 \over 2}, 4$ && $1$                 && ${338 \over 27}$ && ${50 \over 3}$
                          && $-{130 \over 9}$
                            && $n = 3$, $k=1$ &\cr \tablespace
&  \omit           && $-{403 \over 22}$   && $-{25440 \over 35213}$
                              && $-{26733375 \over 3732578}$
                                     && $-{80100 \over 35213}$
                        && $n = 3$, $k=-{13 \over 4}$ &\cr \tablerule
& ${9 \over 2}, 4$ && $1$    && ${578 \over 27}$
                          && ${50 \over 3}$
                           && $-{170 \over 9}$
                            && $n = 4$, $k=1$ &\cr \tablespace
&                  && $-{141 \over 2}$    && ${404984 \over 220113}$
                          && ${173683359 \over 79876562}$
                           && $-{48918 \over 24457}$
                            && $n = 4$, $k=-{28 \over 5}$ &\cr \tablespace
&  \omit           && $-{779 \over 26}$   && ${265856 \over 1206249}$
                              && ${392741325 \over 215516488}$
                                     && $-{254820 \over 402083}$
                        && $n = 4$, $k=-{19 \over 4}$ &\cr \tablerule
& ${7 \over 2}, 6$ && ${561 \over 2}$    && ${102056176 \over 4379914}$
                          && ${1250785008225 \over 19866654531862}$
                            && ${16775775 \over 4379914}$
                              && not rational &\cr \tablerule
& ${9 \over 2}, 6$ && $-{304 \over 5}$   && $-{210945917 \over 96451872}$
                              && $-{12132099470456 \over 90830931184851}$
                                     && $-{3258157 \over 6028242}$
                              && not rational &\cr \tablespace
}
\hrule}\hskip 1pt \vrule}
\hbox{\quad Table 2: $c$-values and coupling constants
            for mixed $\w(2,\delta_1, \delta_2)$-algebras}}
}
\mn
If the algebra $\w(2, {5 \over 2}, 6)$ would exist, it would contain a
$\w(2, {5 \over 2})$-subalgebra which is consistent only for
$c=-{13 \over 14}$. Even for this value of the central charge the Jacobi
identity involving the fermion twice and the boson once cannot be satisfied.
Thus, $\w(2, {5 \over 2}, 6)$ does not exist at all. This observation
suggests that algebras $\w(2, \delta_1, \delta_2)$ with
$\delta_2 \ge 2 \delta_1$, $\delta_1 \ge {5 \over 2}$ do not exist because
otherwise it would have  a fermionic $\w(2,\delta_1)$-subalgebra which are
believed to be classified and none of them admits further extensions.
Furthermore, it is not very plausible that the mixed Jacobi identities can
be satisfied under these conditions.
\mn
The algebra $\w(2,{5 \over 2}, 4)$ was already predicted by coset arguments
$\q{\wattsBn}$ and explicitly realized in terms of five free fermions
$\q{\ahnA} \q{\ahnB}$. We obtain the same structure
constants as in $\q{\ahnA} \q{\ahnB}$. We also see that there are no other
$\w(2, {5 \over 2}, 4)$-algebras besides the one constructed by Ahn.
$\w(2, {5 \over 2}, 4)$ reduces to $\w(2, {5 \over 2})$ at
$c=-{13 \over 14}$ which was the first
well studied non-linear fermionic algebra $\q{\zam} \q{\rva}$.
\sn
There have been earlier studies of $\w(2,{7 \over 2}, 4)$ as well
$\q{\zhanghuang}$. Before studying Jacobi identities involving both
$U$ and $V$ we obtain the same structure constants as in $\q{\zhanghuang}$.
The mixed Jacobi identities, however, restrict $c$ to the two values
given in the above table, thus completing these earlier studies.
\sn
Of course, one would like to identify the above algebras. Candidates
are $N=1$ supersymmetric models as well as subalgebras of
${\cal WB}(n)$, which should be distinguished from the purely bosonic
${\cal WB}_n$-algebras. ${\cal WB}(n)$ algebras can be obtained
either by quantum Hamiltonian reduction of the affine Lie Super algebras
${\cal B}(0,n)^{(1)} = osp(1 \mid 2 n)^{(1)}$ $\q{\ito}$
or by coset constructions using the non-simply laced affine Lie algebra
${\cal B}_n^{(1)}$ $\q{\ito} \q{\watts}$.
The algebras ${\cal WB}(n)$ contain $n$ bosonic fields
of dimension $2, \ldots , 2n$ and one fermionic field of dimension
$n+{1 \over 2}$. The central charge for a level $k$ minimal model is given
by $\q{\ito} \q{\watts}$:
$$c=\left(n+{1 \over 2}\right)
    \left( 1 - {2 n (2n -1) \over (k+2n)(k+2n-1)} \right).
\eqno({\rm 5})$$
$\w(2, {5 \over 2}, 4)$ is known to be identical to ${\cal WB}(2)$
$\q{\ahnA} \q{\ahnB}$. The only candidates for Super Virasoro minimal
models are $\w(2, {7 \over 2}, 4)$ and $\w(2, {9 \over 2}, 4)$
at $c=1$. However, most $c$-values lie
in the minimal series of ${\cal WB}(n)$ where $n=\delta_1 -{1 \over 2}$.
The corresponding levels $k$ are given in the above table. It is therefore
natural to assume that these algebras can be interpreted as subalgebras
of ${\cal WB}(n)$ for these specific values of the central charge.
In order to give further evidence for this assumption we have studied
the HWRs of these algebras examining Jacobi identities.
\mn
\centerline{\vbox{
\hbox{
\vrule \hskip 1pt
\vbox{ \offinterlineskip
\def\tablespace{height2pt&\omit&&\omit&&\omit&&\omit&\cr}
\def\tablerule{ \tablespace
                \noalign{\hrule}
                \tablespace        }
\hrule
\halign{&\vrule#&
  \strut\hskip 4pt\hfil#\hfil\hskip 4pt\cr
\tablespace
\tablespace
& $\delta_1,\delta_2$ && $c$      && {\it relevant HWRs}  && $\tilde{c}$ &\cr
\tablespace
\tablerule
& ${7 \over 2},4$ &&   $1$      &&
  $h \in \left\{ 0, {1 \over 14}, {1 \over 10}, {2 \over 7}, 1 \right \}$
                                        && $1$             & \cr \tablerule
& \omit           && $-{403 \over 22}$ &&
  $h \in \left\{ 0, {6 \over 11}, -{4 \over 11}, -{6 \over 11},
                   -{7 \over 11}, -{9 \over 11},
                   -{5 \over 22}, -{13 \over 22},
                   -{15 \over 22}, -{17 \over 22} \right \}$
                                        && ${29 \over 22}$ & \cr \tablerule
& ${9 \over 2},4$ &&   $1$      &&
  $h \in \left\{ 0, 1, {1 \over 2}, {2 \over 9},{8 \over 9},
                 {1 \over 18}, {1 \over 16}  \right \}$
                                        && $1$             & \cr \tablespace
& \omit           && $-{141 \over 2}$ &&
  $h \in \left\{ 0, -2, -3, -{5 \over 2}, -{8 \over 3},
      -{7 \over 4}, -{9 \over 4}, -{11 \over 4},
      -{7 \over 6}, -{13 \over 6}, -{17 \over 6},
      -{23 \over 12}, -{35 \over 12}  \right \}$
                                        && ${3 \over 2}$   & \cr \tablespace
& \omit           && $-{779 \over 26}$ &&
  $h \in {\bigl\{} 0, {8 \over 13}, -{6 \over 13}, -{10 \over 13},
      -{11 \over 13}, -{12 \over 13}, -{14 \over 13},
      -{15 \over 13}, -{17 \over 13},$  && ${37 \over 26}$ & \cr \tablespace
& \omit           && \omit &&
      $\qquad \qquad
      -{7 \over 26}, -{19 \over 26}, -{27 \over 26},
      -{29 \over 26}, -{31 \over 26}, -{33 \over 26}  {\bigr \}}$
                                        && \omit        & \cr \tablerule
& ${7 \over 2},6$ &&   ${561 \over 2}$      &&
$\vt=- {h (481511 - 88782 h +4096 h^2) \over 67103100} \ , \
\vt=- {23890440 -2256769 h - 216382 h^2 + 20480 h^3 \over 335515500}$
                                        && \omit        & \cr \tablerule
& ${9 \over 2},6$ &&   $-{304 \over 5}$      &&
$\vt= {h (713786 + 564905 h + 111725 h^2) \over 55854120} \ , \
\vt= {6576264 + 6369916 h + 1886911 h^2 + 156415 h^3 \over 78195768}$
                                        && \omit        & \cr \tablespace
}
\hrule}\hskip 1pt \vrule}
\hbox{\quad Table 3: Representations of mixed
                     $\w(2,\delta_1, \delta_2)$-algebras}}
}
\mn
In this table, $\vt$ denotes the eigenvalue of $V_0 (C_{VV}^V)^{-1}$.
A finite set of $h$-values has to be understood in the sense that also
the corresponding values of $\vt$ are fixed.
\sn
In particular for $\w(2, {7 \over 2}, 4)$ and $\w(2, {9 \over 2}, 4)$
at $c=1$ the explicit values
of $h$ show that these models are indeed not Super Virasoro minimal.
These two algebras confirm the expectation that $\w(2,\delta_1,4)$ exists
for $c=1$. This is inferred from the construction of modular invariant
partition functions in $c=1$ theories $\q{\gin}$.
\mn
One of the two exceptions -- $\w(2, {7\over 2},6)$ at $c={561 \over 2}$ --
is certainly not rational because $c$ lies above the bound $c={5 \over 2}$
for rational models with two bosons and one fermion $\q{\wirrep}$.
Our explicit results are in agreement with this prediction.
The result for $\w(2, {9\over 2},6)$ at $c=-{304 \over 5}$ also has
a structure which is typical for $\w$-algebras which are not rational.
\bn
\section{4.\ Representations of $\bf \ww(2, 4, 4)$ and $\bf \ww(2, 4, 5)$}
\mn
In this and the next section we shall examine those bosonic $\w$-algebras
with two additional generators which have been constructed by $\q{\blm}
\q{\kau}$. $\w(2,3,4)$ has been identified as the Casimir algebra
${\cal WA}_3$. For the remaining $\w(2,\delta_1,\delta_2)$-algebras
we explicitly studied HWRs. We denote the additional generators
by $U$ and $V$ where the dimension of $U$ is less or equal
to the dimenion of $V$. The eigenvalues of $U_0$ and $V_0$
will be denoted by $u$ and $v$. Usually it is more convenient to
consider the normalization independent quantities $\ut := u (C_{UU}^U)^{-1}$,
$\vt := v (C_{VV}^V)^{-1}$.
\sn
The results for $\delta_1 = 4$ are listed in the following table:
\mn
\centerline{\vbox{
\hbox{
\vrule \hskip 1pt
\vbox{ \offinterlineskip
\def\tablespace{height2pt&\omit&&\omit&&\omit&&\omit&&\omit&\cr}
\def\tablerule{ \tablespace
                \noalign{\hrule}
                \tablespace        }
\hrule
\halign{&\vrule#&
  \strut\hskip 4pt\hfil#\hfil\hskip 4pt\cr
\tablespace
\tablespace
& $\delta_2$ && $c$ && {\it relevant HWRs}   && $\tilde{c}$
                                             && {\it interpret.} &\cr
\tablespace
\tablerule
& $4$ &&   $1$      && $h \in \{0, 1, {1 \over 4}, {1 \over 16},
                           {9 \over 16} \}$ && $1$ && ${\cal WD}_4$, $c_{7,8}$
                                                & \cr \tablespace
&\omit&&$-{656\over 11}$&&$h \in \{0, -{19 \over 11}, -{26 \over 11},
                           -{27 \over 11}, -{28 \over 11}, -{12 \over 11},
                           -{18 \over 11}, -{21 \over 11}, -{24 \over 11},
                           -{25 \over 11} \}$  && ${16 \over 11}$ &&
                        ${\cal WD}_4$, $c_{6,11}$ & \cr \tablerule
& $5$ &&   $1$      && $h \in \{0, 1, {1 \over 20}, {9 \over 20},
                           {1 \over 5}, {4 \over 5}, {5 \over 4},
                           {1 \over 16}, {9 \over 16} \}$ && $1$
                        && ${\cal WD}_5$, $c_{9, 10}$ & \cr \tablespace
&\omit&&$-{1060\over 13}$&& $h \in \{0, -3, -{16 \over 13}, -{24 \over 13},
                           -{27 \over 13}, -{33 \over 13}, -{34 \over 13},
                           -{35 \over 13}, -{36 \over 13},$ && ${20 \over 13}$
                        &&  ${\cal WD}_5$, $c_{8,13}$ & \cr \tablespace
& \omit && \omit      && $\quad -{38 \over 13}, -{41 \over 13}, -{43 \over 13},
                           -{44 \over 13}, -{45 \over 13}, -{42 \over 13},
                           -{20 \over 13}, -{30 \over 13}, -{32 \over 13}
                                        \}$  && \omit && \omit
                                                        & \cr \tablespace
&\omit&&$-{253\over 7}$&& $\ut=-{h (829 + 812 h) \over 7380}
                  \and v^2 = {32 h^2 (6 + 7 h)^2 (8 + 7 h) \over 1645875}$
                        && \omit && ${\cal WD}_5$, $c_{5,7}$
                                                        & \cr \tablespace
&\omit&&\omit&& $\ut=-{29997+40757 h+13804 h^2\over 125460}
       \and v^2 = {8 (3 + 2 h)^2 (9 + 7 h)^2 (11+ 7 h) \over 1645875}$
                                && \omit   &&  \omit & \cr \tablespace
&\omit&&\omit&& $\ut=-{122210+223083 h+96628 h^2\over 878220}
       \and v^2 = {2 (19 + 14 h)^2 (11 + 14 h)^2 (10+ 7 h) \over 80647875}$
                                && \omit   &&  \omit & \cr \tablespace
}
\hrule}\hskip 1pt \vrule}
\hbox{\quad Table 4: Representations of $\w(2, 4, \delta_2)$-algebras}}
}
\mn
Here, a finite set of $h$-values has to be understood in the sense
that also the eigenvalues $u$, $v$ of the additional simple fields
are fixed to some
discrete values. In the above table this data has not been presented,
neither have the multiplicities of $h$-values been indicated.
\sn
One remarks that the values of the central charge of
$\w(2,4,4)$ and $\w(2,4,5)$ lie in the minimal series of ${\cal WD}_n$
which is given by coprime $p$, $q$ and
$$c_{p,q} = n \left(
1-(2 n -1) (2 n -2) {(p-q)^2 \over p q} \right) \ ,
\qquad p,q \ge 2 (n-1).    \eqno({\rm 6})$$
Although $c=-{253 \over 7}$ can be parametrized for $n=5$ according to
(6), this is not a minimal model because $p=5$ is too small. This
is reflected in our explicit results which are typical for non-rational
models. For all other values of the central charge our explicit
calculations demonstrate rationality. Furthermore, the $h$-values can
be obtained from the general formula presented in $\q{\frenkel}$
for the minimal models of Casimir algebras specialized to ${\cal D}_n$.
This supports the identification of these algebras with (sub-)algebras
of ${\cal WD}_n$ for particular values of $c$ $\q{\kausch}$.
\bn
\section{5.\ The exception: An uninterpreted generic solution
                            for $\bf \ww(2, 4, 6)$}
\mn
Finally, let us focus on the algebra $\w(2,4,6)$ and try to clarify
some open questions about the interpretation of one solution.
For this algebra there exist three solutions $\q{\kauW}$. The first
one has been identified $\q{\horst} \q{\commute}$
with the bosonic projection of the  $N=1$-Super-Virasoro algebra
proposed by $\q{\bouwknegt}$. The existence of the second solution
was not completely clear in $\q{\kau}$ but now it is supposed to yield
the Casimir algebras of ${\cal WB}_3$, ${\cal WC}_3$ $\q{\kauW}$.
The solution we are going to discuss is the third one.
The coupling constants for the simple fields of this third solution
to $\w(2,4,6)$ are determined by $\q{\kau} \q{\kliem}$:
$$\eqalign{
C_{V V}^{U} &= {10 (7 c + 68) (2 c - 1) (c + 50)
          \over 9 (5 c^2 + 309 c - 14) (c + 24) } C_{U U}^{U} \cr
C_{U U}^{V} &= {9 (14 c + 11) (5 c + 22) (c + 76) (c + 24) (c - 1)
           \over 5 (1106 c^5 + 72355 c^4 + 839120 c^3 -
          574940 c^2 - 1961680 c - 2510336)} C_{V V}^{V} \cr
\left( C_{U U}^{U} \right)^2
       &= -{(5 c^2+309 c-14)^2  \over (5 c+22) (5 c+3) (c-26)} \cr
\left( C_{V V}^{V} \right)^2
       &= -{50 (1106 c^5+72355 c^4+839120 c^3-574940 c^2
                    -1961680 c-2510336)^2 (c+49)  \over
                 243 (14 c+11) (7 c+68) (5 c+3) (2 c-1) (c+76)
                       (c+24)^3 (c-1) (c-26)}. \cr
}   \eqno({\rm 7})$$
For the solution (7) we have checked the validity of {\it all}
Jacobi-Identities connecting the simple additional fields.
\smallskip
The solution we are interested in gives rise to a null field
of dimension 11 for generic value of the central charge. This is
something observed so far only for orbifolds of $\w$-algebras.
Using this null field we were able to derive one relation
among the $L_0$-eigenvalue $h$, $u$ and $v$
for all values of the central charge. Thus, at most two
of the three parameters $h$, $u$ and $v$ are in fact independent.
This is a further hint for an orbifold-construction
or something similar. If this solution for $\w(2,4,6)$ should
indeed be an orbifold, at least some of
the simple fields $\phi_i$ in the original algebra
would lead to HWRs with $h_i$ equal to the
dimension of $\phi_i$. A corresponding highest weight
vector is given by $\mid \!h_i \rangle\, = \phi_{i,h_i} \vac$.
Note that this is indeed true for the representations of $\w(2,10)$
at $c=-2$.
\sn
The algebra $\w(2,4,6)$ reduces to $\w(2,4)$ for
$c \in \left\{1, -49, -76, -{11 \over 14} \right\}$ as we can see from (7).
The representations of $\w(2,4)$ have already been
studied in $\q{\wirrep}$. The models at $c=1$ and
$c=-76$ were shown to be non-rational, $c=-49$ is not accessible
to the null field method leaving $c= -{11 \over 14}$ as the
only rational model among them. Here, only $h={3 \over 2}$
lies in ${\zed_{+} \over 2}$ and thus is the only candidate for
a field that has been projected out. Note however that at
$c= -{11 \over 14}$ the bosonic projection of the $N=1$-Super
Virasoro algebra $\q{\commute}$ and the solution (7)
for $\w(2,4,6)$ reduce to the same $\w(2,4)$ and we just see
remnants of the orbifolding in the first solution.
Therefore it is important to study as many additional rational
models for $\w(2,4,6)$ as possible.
\sn
For $c=-{13 \over 15}$ there exists a second null
field with dimension 8. With the aid of this null field we were
able to derive two further relations among $h$, $u$ and $v$
such that we obtained a rational model. At $c=-{10 \over 11}$
a  null field with dimension 10 exists which also
enabled us to derive a rational model. We list the $h$-values of
these two rational models in the following table:
\mn
\centerline{\vbox{
\hbox{
\vrule \hskip 1pt
\vbox{ \offinterlineskip
\def\tablespace{height2pt&\omit&&\omit&&\omit&\cr}
\def\tablerule{ \tablespace
                \noalign{\hrule}
                \tablespace        }
\hrule
\halign{&\vrule#&
  \strut\hskip 4pt\hfil#\hfil\hskip 4pt\cr
\tablespace
\tablespace
& $c$      && $h$-{\it values} && $\tilde{c}$ &\cr
\tablespace
\tablerule
& $-{13 \over 15}$ &&
   $0, {1 \over 3}, {2 \over 3}, {9 \over 4}, {7 \over 5},
    {5 \over 9}, {19 \over 12}, -{1 \over 12}, {11 \over 15},
    {1 \over 15}, {13 \over 20}, {29 \over 36}, -{2 \over 45},
    {19 \over 60}, -{1 \over 60}, {37 \over 180}$
&& ${17 \over 15}$ & \cr \tablerule
& $-{10 \over 11}$ &&
   $0, 3, {4 \over 3}, {5 \over 8}, {17 \over 8},
    -{1 \over 11}, {4 \over 11}, {7 \over 11}, {9 \over 11},
    {10 \over 11}, {18 \over 11}, {26 \over 11},
    -{1 \over 33}, {5 \over 33}, {8 \over 33}, {23 \over 33},$
&& ${14 \over 11}$ & \cr \tablespace
& \omit &&
   $-{5 \over 88}, -{1 \over 88}, {3 \over 88}, {23 \over 88},
    {39 \over 88}, {47 \over 88}, {67 \over 88}, {131 \over 88}$
&& \omit           & \cr \tablespace
}
\hrule}\hskip 1pt \vrule}
\hbox{\quad Table 5: $h$-values for rational models of $\w(2, 4, 6)$}}
}
\mn
Most of the $h$-values have unique values for $u$ and $v$, i.e.\
a multiplicity of at most one. The only exception is
$h={9 \over 11}$ at $c=-{10 \over 11}$
which admits two different values for $u$, $v$ thus leading
to a possible multiplicity of 2.
Note that $c=-{13 \over 15}$ appears in the minimal series of ${\cal WB}_3/
{\cal WC}_3$ and $c=-{10 \over 11}$ in the minimal series of ${\cal WB}_4/
{\cal WC}_4$.
\sn
At $c=-50$ and $c=-{248 \over 5}$ there are also null fields
of dimension 8 resp.\ 10 but explicit calculations indicated
non-rational models with several one-parameter branches of
representations. We omit the explicit results for these models.
\sn
Clearly, at $c=-{13 \over 15}$ there are
neither integer nor half-integer
dimensions such that we have to give up the idea of a
usual orbifold. Furthermore, it is neither significant that
in both models presented in Table 5 there are some
representations with $h \in {\zed \over 3}$ because this
is not so for $c = -{11 \over 14}$.
\medskip
One might expect that a classical analogue for this
exceptional solution exists. However, it was already
pointed out in $\q{\bowwatts}$ that this solution
as well as the orbifold of the $N=1$ Super Virasoro algebra
cannot be obtained via a standard limiting procedure due
to divergent structure constants.
The rescaling $U' = c^{-{1 \over 2}} U$,
$V' = c^{-{1 \over 2}} V$ might lead to a well-defined
classical algebra but in the limit the commutators of $U'$
and $V'$ will not contain any central term.
It is also interesting to consider the limit
$c \to \infty$ of the vacuum preserving algebra (VPA)
$\q{\bowwatts}$ for these two solutions. It is
well-defined in the basis of the rescaled fields $L$,
$U'$ and $V'$. Some of the normal
ordered products containing an even number of simple
fields $U'$, $V'$ (e.g.\ $\n(U',U')$ ) do not decouple
in the limit. Thus, it does not lead to a finite
Lie algebra -- in contrast to the limits of VPAs of
Casimir algebras $\q{\bowwatts}$.
\medskip
In summary, our explicit results for the solution (7)
show some features of an orbifold
which clearly conflict with a construction similar
to Casimir-algebras. Even more, our studies of the
HWRs also rule out the possibility of an orbifold
construction of a different generically existing
$\w$-algebra.
\bn
\section{6.\ Conclusion}
\mn
Our study of $\w(2,\delta)$-algebras did not lead to any algebra
that was not covered by a known series -- except that
we found the first example of an orbifold of a non-unitary
$(1,s)$-algebra. It is remarkable that sporadic non-rational
cases known so far $\q{\blm} \q{\wirrep}$ did not turn up
any more for $8 < \delta \le 11$: Neither did we find an algebra
existing for irrational values of the central charge nor did
models with several one parameter branches of HWRs at
rational values of $c$ turn up (compare $\w(2,8)$ e.g.\ at
$c=-{1015 \over 2}$). Thus, there is a good chance that all series
and exceptional cases presented in the introduction do really cover
all models which are rational with respect to some
$\w(2, \delta)$-algebra.
\sn
One can explain all rational models of mixed algebras with three
generators and the rational models of $\w(2,4,4)$ and $\w(2,4,5)$
as subalgebras or contractions of the Casimir
algebras ${\cal WB}(n)$, ${\cal WD}_n$. However, we showed that
one solution for $\w(2,4,6)$ is neither a Casimir algebra nor an
orbifold of a different generically existing algebra.
Thus, it has become even more
mysterious why this solution exists (there are no doubts about its
existence any more). A similar phenomenon has been observed
for $\w(2,3,4,5)$ $\q{\hornfeck}$ where also generic null fields
appear and a possible candidate for an orbifold
construction is not known. In order to define a classical limit
a similar rescaling has to be performed $\q{\hornfeck}$ and
also the behaviour of the VPA in the
limit $c \to \infty$ is identical to that of the unidentified
solution to $\w(2,4,6)$. Thus, these unexplained algebras may
have a common origin, or even belong to a large class of yet
unknown algebras.
\sn
A possibly related problem turns up in the classification of
$c = 24$ conformal field theories with a single character
$\q{\schellekens}$. Although 71 models of this type are
admitted $\q{\schellekens}$ only 40 of them have been
constructed explicitly up to now $\q{\flm - \dgm}$.
Most constructions start from Niemeier lattices
which yield symmetry algebras that consist of
abelian currents and vertex operators with conformal
dimensions greater or equal to one $\q{\dgm} \q{\schellekens}$.
Other theories are more complicated,
e.g.\ the symmetry algebra of the one invariant under the
Monster group is a $\w(2^{196884}, 3^{21296876})$ --
a surprisingly small algebra compared to the order of
the Monster. For the construction of
the remaining 31 theories one probably needs new methods
that could be related to unexplained $\w$-algebras like
$\w(2,4,6)$.
\sn
In summary, we have shown that quantum $\w$-algebras exist which
cannot be explained as the analogue of a classical $\w$-algebra
\footnote{${}^{2})$}{\sevenrm
For example Casimir algebras have classical
analoga.} or subalgebra respective orbifold thereof.
Certainly, exceptional $\w$-algebras
have to be better understood --in particular the yet uninterpreted
solution for $\w(2,4,6)$-- and exotic constructions on them
have to be studied (compare $\w(2,8)$ at $c=-{712 \over 7}$
$\q{\wirrep}$)
before one can use $\w$-algebras for the classification of
RCFTs.
\bn
\section{Acknowledgements}
\mn
It is a pleasure to thank R.\ Blumenhagen, M.\ Flohr, G.\ von Gehlen,
H.G.\ Kausch,
W.\ Nahm, A.\ Recknagel, R.\ Varnhagen and everybody working in the
theory group of the Physikalisches Institut Bonn for
helpful discussions on various related subjects.
\sn
We are indebted to the Max-Planck-Institut f\"ur Mathematik, Bonn-Beuel
(MPIM) because a great part of the results in this letter has been
obtained with the aid of their computers.
\sn
W.E.\ is grateful for financial support by the MPIM. \par\noindent
R.H.\ would like to thank the
NRW-Graduiertenf\"orderung for a research studentship.
\vfill
\eject
\section{References}
\sn
\settabs\+&\phantom{---------}&\phantom{
------------------------------------------------------------------------------}
& \cr
\+ &$\q{\bpz}$ & A.A.\ Belavin, A.M.\ Polyakov, A.B.\ Zamolodchikov & \cr
\+ &           & {\it Infinite Conformal Symmetry in Two-Dimensional Quantum
                   Field Theory}  & \cr
\+ &           & Nucl.\ Phys.\ {\bf B241} (1984) p.\ 333  & \cr
\+ &$\q{\zam}$ & A.B.\ Zamolodchikov & \cr
\+ &           & {\it Infinite Additional Symmetries in Two-Dimensional
                   Conformal Quantum Field}  & \cr
\+ &           & {\it Theory}, Theor.\ Math.\ Phys.\ 65 (1986) p.\ 1205  & \cr
\+ &$\q{\bbss}$
               & F.A.\ Bais, P.\ Bouwknegt, M.\ Surridge, K.\ Schoutens & \cr
\+ &           & {\it Extensions of the Virasoro Algebra Constructed
                  from Kac-Moody Algebras Using} & \cr
\+ &           & {\it Higher Order Casimir Invariants},
                  Nucl.\ Phys.\ {\bf B304} (1988) p.\ 348 & \cr
\+ &$\q{\blg}$ & A.\ Bilal, J.L.\ Gervais & \cr
\+ &           & {\it Systematic Construction of Conformal Theories with
                   Higher-Spin Virasoro} & \cr
\+ &           & {\it Symmetries}, Nucl.\ Phys.\ {\bf B318} (1989) p.\ 579 &\cr
\+ & $\q{\drinso}$
               & V.G.\ Drinfeld, V.V.\ Sokolov, {\it Lie Algebras and Equations
                  of Korteweg-de Vries Type} & \cr
\+ &           & Jour.\ Sov.\ Math.\ {\bf 30} (1985) p.\ 1975 & \cr
\+ & $\q{\feifre}$
               & B.\ Feigin, E.\ Frenkel, {\it Quantization of the
                   Drinfeld-Sokolov Reduction} & \cr
\+ &           & Phys.\ Lett.\ {\bf B246} (1990) p.\ 75 & \cr
\+ & $\q{\bouwschou}$
               & P.\ Bouwknegt, K. Schoutens, {\it $\w$-Symmetry in Conformal
                   Field Theory}  & \cr
\+ &           & Phys.\ Rep.\ {\bf 223} (1993) p.\ 183 & \cr
\+ & $\q{\laszlo}$
               & L.\ Feh\'er,  L.\ O'Raifeartaigh, P.\ Ruelle,
                   I.\ Tsutsui, A.\ Wipf & \cr
\+ &           & {\it On Hamiltonian Reductions of the
                   Wess-Zumino-Novikov-Witten Theories} & \cr
\+ &           & Phys.\ Rep.\ {\bf 222} (1992) p.\ 1 & \cr
\+ & $\q{\ragoucy}$
               & L.\ Frappat, E.\ Ragoucy, P.\ Sorba,
                 {\it $\w$-Algebras and Superalgebras from Constrained} & \cr
\+ &           & {\it WZW Models: A Group Theoretical Classification} & \cr
\+ &           & preprint ENSLAPP-AL-391/92, hep-th/9207102 & \cr
\+ &$\q{\blm}$ & R.\ Blumenhagen, M.\ Flohr, A.\ Kliem,
                    W.\ Nahm, A.\ Recknagel, R.\ Varnhagen & \cr
\+ &           & {\it $\w$-Algebras with Two and Three Generators},
                    Nucl.\ Phys.\ {\bf B361} (1991) p.\ 255 & \cr
\+ &$\q{\kau}$ & H.G.\ Kausch, G.M.T.\ Watts, {\it A Study of $\w$-Algebras
                   Using Jacobi Identities}& \cr
\+ &           & Nucl.\ Phys.\ {\bf B354} (1991) p.\ 740 & \cr
\+ &$\q{\mfl}$ & M.\ Flohr, {\it $\w$-Algebras,
                   New Rational Models and Completeness of the $c=1$} & \cr
\+ &           & {\it Classification}, preprint BONN-HE-92-08 (1992),
                   to be published in Comm.\ Math.\ Phys.\ & \cr
\+ &$\q{\rva}$ & R.\ Varnhagen, {\it Characters and Representations of
                   New Fermionic $\w$-Algebras} & \cr
\+ &           & Phys.\ Lett.\ {\bf B275} (1992) p.\ 87 & \cr
\+ &$\q{\wirrep}$
               & W.\ Eholzer, M.\ Flohr, A.\ Honecker,
                   R.\ H{\"u}bel, W.\ Nahm, R.\ Varnhagen  & \cr
\+ &           & {\it Representations of $\w$-Algebras with Two Generators
                   and New Rational Models } & \cr
\+ &           & Nucl.\ Phys.\ {\bf B383} (1992) p.\ 249 & \cr
\+ &$\q{\cap}$ & A.\ Cappelli, C.\ Itzykson, J.B.\ Zuber & \cr
\+ &           & {\it The A-D-E Classification of Minimal and $A_1^{(1)}$
                  Conformal Invariant Theories} & \cr
\+ &           & Comm.\ Math.\ Phys.\ 113 (1987) p.\ 1 & \cr
\+ &$\q{\wowoA}$
               & W.\ Eholzer, {\it Fusion Algebras Induced by Representations
                   of the Modular Group} & \cr
\+ &           & Int.\ Jour.\ of Mod.\ Phys.\ {\bf A8} (1993) p.\ 3495 & \cr
\+ &$\q{\wowoB}$
               & W.\ Eholzer, N.\ Skoruppa, {\it Exceptional
                  $\w$-Algebra Characters and Theta-Series} & \cr
\+ &           & {\it of Quaternion Algebras}, in preparation & \cr
\+ &$\q{\gin}$ & P.\ Ginsparg, {\it Curiosities at c=1},
                 Nucl.\ Phys.\ {\bf B295} (1988) p.\ 153 & \cr
\+ &$\q{\kri}$ & E.B.\ Kiritsis, {\it Proof of the Completeness
                 of the Classification of Rational Conformal} & \cr
\+ &           & {\it Theories with c=1},
                 Phys.\ Lett.\ {\bf B217} (1989) p.\ 427 & \cr
\vfill
\eject
\+ &$\q{\kausch}$
               & H.G.\ Kausch & \cr
\+ &           & {\it Extended Conformal Algebras Generated by a Multiplet
                   of Primary Fields} & \cr
\+ &           & Phys.\ Lett.\ {\bf B259} (1991) p.\ 448 & \cr
\+ & $\q{\flohkau}$
               & M.\ Flohr, H.G.\ Kausch,
                 {\it A New Series of Rational Conformal Field Theories} & \cr
\+ &           & in preparation & \cr
\+ & $\q{\hornfeck}$
               & K.\ Hornfeck, {\it $\w$-Algebras with Set of
                   Primary Fields of Dimensions}  & \cr
\+ &           & {\it (3,4,5) and (3,4,5,6)},
                 preprint KCL-TH-92-9, DFTT-70/92, hep-th/9212104  & \cr
\+ &           & to be published in Nucl.\ Phys.\ {\bf B} & \cr
\+ & $\q{\bouwknegt}$
               & P.\ Bouwknegt, {\it Extended Conformal Algebras from
                 Kac-Moody Algebras} & \cr
\+ &           & Proceedings of the meeting `Infinite dimensional
                 Lie algebras and Groups' & \cr
\+ &           & CIRM, Luminy, Marseille (1988) p.\ 527 & \cr
\+ &$\q{\kauW}$
               & H.G.\ Kausch, private communication \cr
\+ & $\q{\commute} $
               & A.\ Honecker, {\it A Note on the Algebraic Evaluation
                    of Correlators in Local Chiral} & \cr
\+ &           & {\it Conformal Field Theory},
                    preprint BONN-HE-92-25 (1992), hep-th/9209029 & \cr
\+ &$\q{\kliem}$
               & A.\ Kliem, {\it Konstruktion von $\w$-Algebren},
                  Diplomarbeit BONN-IR-91-46 (1991) &\cr
\+ & $\q{\twists}$
               & A.\ Honecker & \cr
\+ &           & {\it Automorphisms of $\w$-Algebras
                 and Extended Rational Conformal Field Theories}& \cr
\+ &           & Nucl.\ Phys.\ {\bf B400} (1993) p.\ 574 & \cr
\+ &$\q{\wattsBn}$
               & G.M.T.\ Watts, {\it $\w$-Algebras and Coset Models},
                 Phys.\ Lett.\ {\bf B245} (1990) p.\ 65 & \cr
\+ &$\q{\ahnA}$
               & C.\ Ahn, {\it $c={5 \over 2}$ Free Fermion Model of
                                ${\cal WB}_2$ Algebra} \cr
\+ &           & Int.\ Jour.\ of Mod.\ Phys.\ {\bf A7} (1992) p.\ 6799 & \cr
\+ &$\q{\ahnB}$
               & C.\ Ahn, {\it Explicit Construction of Spin 4 Casimir
                               Operator in the Coset Model} \cr
\+ &           & {\it $\hat{SO}(5)_1\times \hat{SO}(5)_m/\hat{SO}(5)_{1+m}$},
                 preprint ITP.SB-92-45, hep-th/9209001  & \cr
\+ &$\q{\zhanghuang}$
               & D.H.\ Zhang, C.S.\ Huang, {\it Spin-4 and -${7 \over 2}$
                          Extended Conformal Algebra} \cr
\+ &           & Europhys.\ Lett.\ {\bf 14} (1991) p.\ 313 & \cr
\+ &$\q{\ito}$
               & K.\ Ito, {\it Quantum Hamiltonian Reduction and ${\cal WB}$
                               Algebra} \cr
\+ &           & Int.\ Jour.\ of Mod.\ Phys.\ {\bf A7} (1992) p.\ 4885 & \cr
\+ &$\q{\watts}$
               & G.M.T.\ Watts,
                 {\it ${\cal WB}$ Algebra Representation Theory},
                 Nucl.\ Phys.\ {\bf B339} (1990) p.\ 177 & \cr
\+ & $\q{\frenkel}$
               & E.\ Frenkel, V.\ Kac, M.\ Wakimoto & \cr
\+ &           & {\it Characters and Fusion Rules for $\w$-Algebras
                    via Quantized Drinfeld-Sokolov Reduction} & \cr
\+ &           & Comm.\ Math.\ Phys.\ 147 (1992) p.\ 295 & \cr
\+ & $\q{\horst}$
               & H.G.\ Kausch, {\it Chiral Algebras in Conformal Field
                     Theory} & \cr
\+ &           & Ph.D.\ thesis, Cambridge University, September 1991 & \cr
\+ &$\q{\bowwatts}$
               & P.\ Bowcock, G.M.T.\ Watts,
                 {\it On the Classification of Quantum $\w$-Algebras}& \cr
\+ &           & Nucl.\ Phys.\ {\bf B379} (1992) p.\ 63 & \cr
\+ & $\q{\schellekens}$
               & A.N.\ Schellekens, {\it Meromorphic $c = 24$ Conformal
                 Field Theories} & \cr
\+ &           & Comm.\ Math.\ Phys.\ 153 (1993) p.\ 159 & \cr
\+ & $\q{\flm}$& I.B.\ Frenkel, J.\ Lepowsky, A.\ Meurman,
                 {\it A Moonshine Module for the Monster} & \cr
\+ &           & Vertex Operators in Mathematics and Physics, & \cr
\+ &           & eds.\ J.\ Lepowsky, S.\ Mandelstam, I.M.\ Singer,
                 Springer Verlag (1985) p.\ 231 & \cr
\+ & $\q{\tuite}$
               & M.P.\ Tuite, {\it Monstrous Moonshine from Orbifolds} & \cr
\+ &           & Comm.\ Math.\ Phys.\ 146 (1992) p.\ 277 & \cr
\+ & $\q{\dgm}$& L.\ Dolan, P.\ Goddard, P.\ Montague & \cr
\+ &           & {\it Conformal Field Theory, Triality and the Monster
                 Group} & \cr
\+ &           & Phys.\ Lett.\ {\bf B236} (1990) p.\ 165 & \cr
\vfill
\eject
\end